\documentclass[final,3p,times]{elsarticle}

\usepackage[rightcaption]{sidecap}
\usepackage[normalem]{ulem}
\usepackage{amssymb,comment}
\usepackage{amsmath,graphicx,subfigure,float}

\usepackage{xcolor}
\journal{Journal of Subatomic Particles and Cosmology}

\begin{document}

\begin{frontmatter}



\title{QCD Critical Point: Theoretical Constraints and Modeling of Fluctuation Observables}

\author[aaa]{Maneesha Sushama Pradeep}
\ead{maneeshas1@iisc.ac.in}
\affiliation[aaa]{organization={Indian Institute of Science},
             addressline={C.V. Raman Avenue Road},
             city={Bangalore},
             postcode={560012},
             state={Karnataka},
             country={India}}

 \begin{abstract}
This review summarizes recent theoretical progress on the QCD equation of state near the critical point and developments in the maximum entropy freeze-out framework, which provides a systematic connection between hydrodynamic fluctuations and hadronic multiplicity cumulants. Recent applications of this framework, together with advances in the dynamical evolution of critical fluctuations, are also discussed.
\end{abstract}

\begin{keyword}
QCD critical point, fluctuations, equation of state, maximum entropy, factorial cumulants

\end{keyword}

\end{frontmatter}

\section{Introduction}
\label{sec1}

A central open question in Quantum Chromodynamics (QCD) at finite temperature and baryon density is whether the QCD phase diagram has a critical point at large baryon chemical potential. While lattice QCD has established that the transition from hadronic matter to the quark gluon plasma is a smooth crossover at vanishing baryon chemical potential, a variety of theoretical approaches, including effective models, lattice QCD extrapolations, and functional methods, allow for the possible existence of a critical point at finite baryon density, beyond which the transition becomes first order. One of the primary objectives of the Beam Energy Scan (BES) program at the Relativistic Heavy Ion Collider (RHIC) is to search for experimental signatures of this critical point, should it exist, by colliding heavy-ions at different center of mass energies.

\begin{figure}[h]
    \centering    \includegraphics[width=1.0\textwidth]{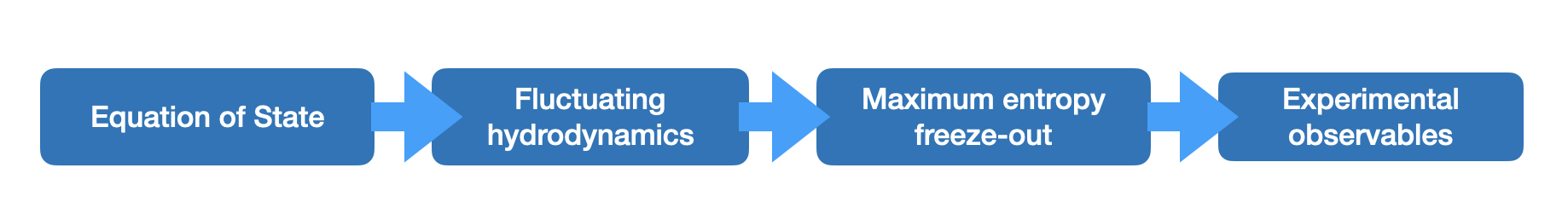}
 \caption{Representation of theoretical pipeline connecting the EoS to experimental observables}
    \label{Fig:pipeline}
\end{figure}

This review summarizes the major components of the theoretical pipeline that is crucial for connecting the QCD Equation of State (EoS) near the critical point to experimental observables, as represented in the Fig.~(\ref{Fig:pipeline}). Sec.~(\ref{Sec:CP}) discusses the bounds and estimates on the location of critical point, and the EoS near its vicinity. Sec.~(\ref{Sec:ME}) reviews recent developments in the maximum entropy freeze-out framework, which provides a systematic procedure for relating hydrodynamic fluctuations to experimentally measurable hadronic fluctuations based on the principle of maximum entropy. Sec.~(\ref{Sec:Dyn}) is devoted to recent theoretical advances in describing the out-of-equilibrium evolution of critical hydrodynamic fluctuations. Finally, Sec.~(\ref{Sec:Sum}) summarizes the main developments with an outlook.

\section{Location of critical point and EoS near its vicinity}
\label{Sec:CP}

The first element of the pipeline in Fig.~\ref{Fig:pipeline} is the EoS. Before fluctuation observables can be predicted, one needs to know where the critical point may lie and how the EoS behaves in its vicinity. The current bounds/estimates on the location of the critical point from lattice and functional approaches to QCD are discussed below.

\subsection{Bounds and estimates on the location of critical point}
    Lattice QCD computations at finite baryon chemical potentials are severely limited due to the infamous sign problem. Nevertheless, they have played an important role in placing stringent constraints on the possible location of the QCD critical point. It is now well established that at vanishing chemical potentials, the chiral cross-over at physical quark masses occurs at a temperature $T^{\text{phys}}_{\text{pc}}=156.5 \, \pm 1.5\,\text{MeV}$\cite{HotQCD:2018pds},
    while the chiral phase transition in the massless quark limit happens at $T^{0}_c\approx 134^{+6}_{-4}\, \text{MeV}$ \cite{Kotov:2021rah}.
    Together with arguments based on the curvature of the $O(4)$ and $Z_2$ critical lines near the tricritical point, which imply the ordering \cite{Hatta:2002sj}
\begin{eqnarray}
T_{\mathrm{pc}}^{\mathrm{phys}} > T_c^{0} > T_{\mathrm{tcp}}^{0} > T_{\mathrm{cp}}^{\mathrm{phys}},
\end{eqnarray}
this hierarchy indicates that the critical point at finite quark masses, if it exists, must lie below the chiral transition temperature in the massless limit, i.e., $T_{\mathrm{cp}}^{\mathrm{phys}} < 134\,\mathrm{MeV}$. In practice, this leads to a conservative lattice upper bound of approximately $T_{\mathrm{cp}} \lesssim 125\,\mathrm{MeV}$ for $\mu_B/T \approx 2$\cite{Ding:2024sux}. Using an improved continuum-extrapolated lattice equation of state with significantly reduced uncertainties, constant-entropy contours were extrapolated from imaginary to real baryon chemical potential.  Assuming that a first-order phase transition would be accompanied by a multivalued entropy density, a critical endpoint for $\mu_B<450\,\mathrm{MeV}$ was excluded at the $2\sigma$ confidence level\cite{Borsanyi:2025dyp}.

The current functional estimate for a chiral critical point at physical quark masses, obtained using a combination of the functional Renormalization Group (fRG) and Dyson-Schwinger (DSE) studies is given by an interval $T(\mu_B)$ of the chiral transition line with a small width,
\begin{eqnarray}
    (T,\mu_B)_{\text{cp}}=(115-105,600-650)\, \text{MeV}.
\end{eqnarray}
For a comprehensive discussion on the phase diagram of QCD obtained from studies based on functional approaches, and quantitative comparisons to lattice calculations, refer to the recent review\cite{Fischer:2026uni}.

\subsection{EoS near the critical point}
\label{Subsec:EoS}
    Knowing where the critical point may lie is not sufficient by itself: predicting fluctuation observables also requires the EoS in its vicinity, including the strength and extent of the critical region. The conjectured critical point of QCD, if it exists, would belong to the 3D Ising universality class. By the universality of critical phenomena, the leading singular part of the EoS near this critical point can be mapped to the Gibbs free energy of the 3D Ising model via a linear mapping between the QCD thermodynamic variables and the Ising reduced temperature and magnetic field. The non-universal mapping parameters determine the location of the critical point ($T_c,\mu_{B,c}$), the orientation of Ising axes on the QCD phase diagram, (via angles $\alpha_1,\alpha_2$) and the strength and size of the critical region in QCD (via scale factors $\rho$ and $w$). A 4D $T'$-expanded lattice QCD equation of state \cite{Abuali:2025tbd} supplemented by a multi-dimensional critical contribution, which extends the critical point on the $(T,\mu_B)$ plane to a critical surface at finite $\mu_B,\mu_Q$ and $\mu_S$ was presented at SQM 2026\cite{JahanSQM}.

   Since the mapping parameters are not fixed by universality, constraining them is essential for quantitative predictions. One route to such constraints is provided by the analytic structure of the EoS in the crossover region. In the crossover region of the QCD phase diagram, the Lee-Yang edge singularity can be extracted from the equation of state reconstructed either from Taylor expansions about $\mu_B=0$ or from calculations at imaginary chemical potential. The universality of the approach of the Lee-Yang singularities to the real axis, in the vicinity of the critical point can be used to bound and constrain the non-universal mapping parameters and thereby, determine the EoS in the vicinity of the critical point. Using improved resummation techniques that blend in Pade approximants and conformal maps, the critical temperature, the critical baryon chemical potential and the slope of the first-order curve at the critical point were estimated as $97^{+18}_{-18}\, \text{MeV}$, $579^{+172}_{-160}\, \text{MeV}$ and $9.40^{\circ\,3.89}_{\,\,-3.81}$ respectively \cite{Basar:2023nkp}. For fixed values of $\mu_c,T_c,\alpha_1$ and $\alpha_2$, a specific combination of $\bar{\rho}=\rho w^{1-\frac{1}{\beta\delta}}$ (where $\beta\approx 0.326$ and $\delta\approx 4.8$ are the critical exponents for 3D Ising universality class) can also be constrained using the same analysis\cite{Basar:2026irk}.

Together, these complementary approaches progressively constrain the critical part of the QCD EoS. The  cumulants of the conserved charge densities in equilibrium are directly related to derivatives of the pressure (EoS) with respect to temperature and the corresponding chemical potentials. Near QCD critical point, the higher order derivatives of pressure diverge as power laws with exponents specific to the 3D Ising universality class. Consequently, the associated thermodynamic fluctuations are strongly enhanced and can give rise to observable consequences, most notably a non-monotonic dependence of hadronic multiplicity cumulants on the collision energy in heavy-ion collisions~\cite{Stephanov:2008qz}. Next section discusses how these enhanced fluctuations translate into phase space correlations of particle multiplicities.

\section{Connecting hydrodynamic fluctuations to event-by-event particle multiplicity fluctuations}
\label{Sec:ME}
\subsection{Maximum Entropy Freeze-out framework}
\label{SubSec:ME}
The enhanced fluctuations discussed above are properties of the hydrodynamic densities, whereas experiments measure event-by-event fluctuations of particle multiplicities. Bridging this gap is the second element of the pipeline. Freezeout (particlization) provides the crucial link between hydrodynamic fluctuations  and phase space
correlations of particle multiplicities observed in experiment. 
For single particle observables, such as mean particle yields, Cooper-Frye prescription has been successful in connecting to the observations. For critical point searches, the observables of interest are event-by-event fluctuations of particle multiplicities. The maximum entropy freeze-out prescription introduced in \cite{Pradeep:2022eil} generalizes the Cooper-Frye prescription by maximizing the entropy of the fluctuations of hadron resonance gas distribution at freeze-out while ensuring that the correlations of conserved densities in the gas ensemble match those of the hydrodynamic description at freeze-out. 

The entropy of the fluctuating hadron resonance gas is mathematically similar to the n-PI action of quantum field theory. Maximizing this entropy, subject to the constraint that the correlations of conserved densities in the hadronic gas reproduce the corresponding hydrodynamic correlations at freeze-out, leads to the compact relation:
\begin{eqnarray}
    \widehat{\Delta} G=\widehat{\Delta} H \cdot P\cdot \dots P
\end{eqnarray}
Here, $\widehat{\Delta}G$ and $\widehat{\Delta}H$ denote the irreducible relative cumulants (IRCs) of particle multiplicities and hydrodynamic densities, respectively, while $P$ is a projection kernel determined entirely by the hadron resonance gas in equilibrium. 
The IRCs quantify the genuinely irreducible out-of-equilibrium correlations that cannot be constructed from lower-order cumulants. Further details of the formalism can be found in Ref.~\cite{Pradeep:2022eil}. After integrating over the relevant phase space acceptance, and normalizing by the corresponding mean multiplicity of the proton, the normalized $n^{th}$ factorial cumulant of the proton multiplicity, denoted by $\widehat{\Delta}\omega_{np}$ is obtained.

The $\widehat{\Delta}H$s are obtained from the correlation functions of hydrodynamic densities, which are obtained from a stochastic hydrodynamic simulation (or a deterministic evolution of hydrodynamic correlation, such as Hydro+\cite{Stephanov:2017ghc}), at freeze-out. In the limit of rapid local equilibration, the hydrodynamic cumulants are determined directly by derivatives of the equation of state. All numerical studies reviewed in this Section adopt this approximation. The principal limitation of these studies is the neglect of critical slowing down, which will be discussed in the next section.

\noindent
\begin{figure}[H]
\begin{minipage}[h]{0.50\textwidth}
\vspace{0pt}
\subsection{Sensitivity to mapping parameters}
\label{SubSec:ME}
\vspace{0.5cm}
With the freeze-out prescription in place, the critical EoS of Sec.~\ref{Subsec:EoS} can be translated into predictions for proton factorial cumulants. 
In Ref.~\cite{Karthein:2025hvl}, the authors investigated the qualitative and quantitative sensitivity of the proton-multiplicity fluctuations to the non-universal mapping parameters 
in the critical EoS. A key finding was that the position of the peaks in the proton factorial cumulants along the freeze-out trajectory is primarily controlled by the combination $\bar{\rho}=\rho w^{1-\frac{1}{\beta\delta}}$ (See Fig.~\ref{fig:Senscombined}), whereas the magnitude of their maximum value is more sensitive to $w$, scaling as $w^{-1-1/\delta}$. The authors also examined a family of freeze-out trajectories displaced from the chiral crossover line by a temperature difference $\Delta T_f$, and found that the maximum of the $k^{th}$ order proton factorial cumulant exhibits the universal scaling behavior $\Delta T_f^{1+1/\delta-k}$. 
It was also shown that the qualitative features of the critical signatures are unchanged by the inclusion of the decay protons. 
\end{minipage}\hfill
\begin{minipage}[h]{0.45\textwidth}
    \centering    
\includegraphics[width=0.7\linewidth]{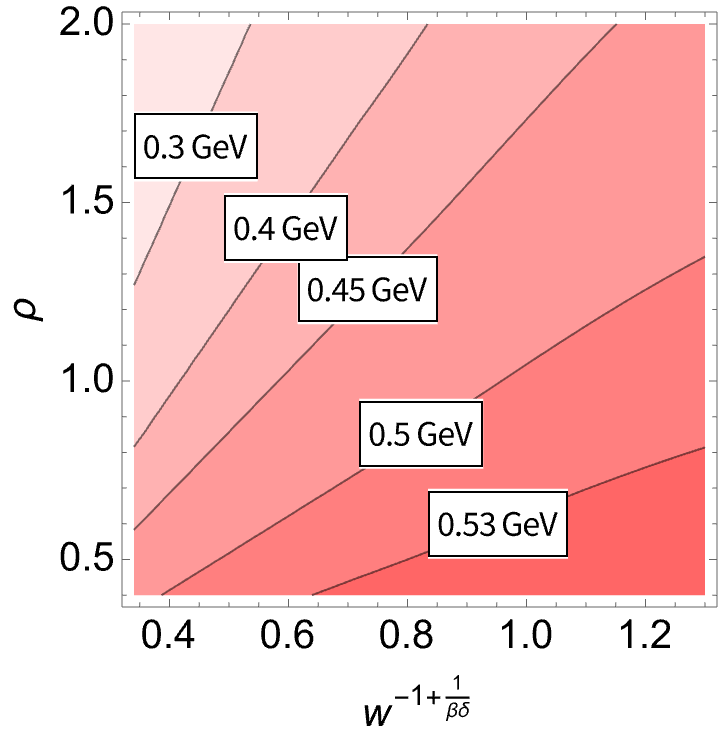}
    \caption{Contours of the baryon chemical potential
  at which the second cumulant of the proton multiplicity, $\hat{\Delta}\omega_{2p}$ along the freezeout curve with $\Delta T_f=6$~MeV, reaches its peak value.}
  \label{fig:Senscombined}
\end{minipage}
  \end{figure}

\subsection{Lee-Yang constrained scenarios}

The study above treated the mapping parameters as free. Since the Lee-Yang analysis of Sec.~\ref{Subsec:EoS} constrains $\bar{\rho}$, the very combination that controls the peak positions, 
the locations of the peaks and dips of the factorial cumulants as a function of collision energy are significantly constrained for a given specification of $\mu_c, T_c$~\cite{Basar:2026irk}.

\begin{figure}[h]
    \centering    \includegraphics[width=1.0\textwidth]{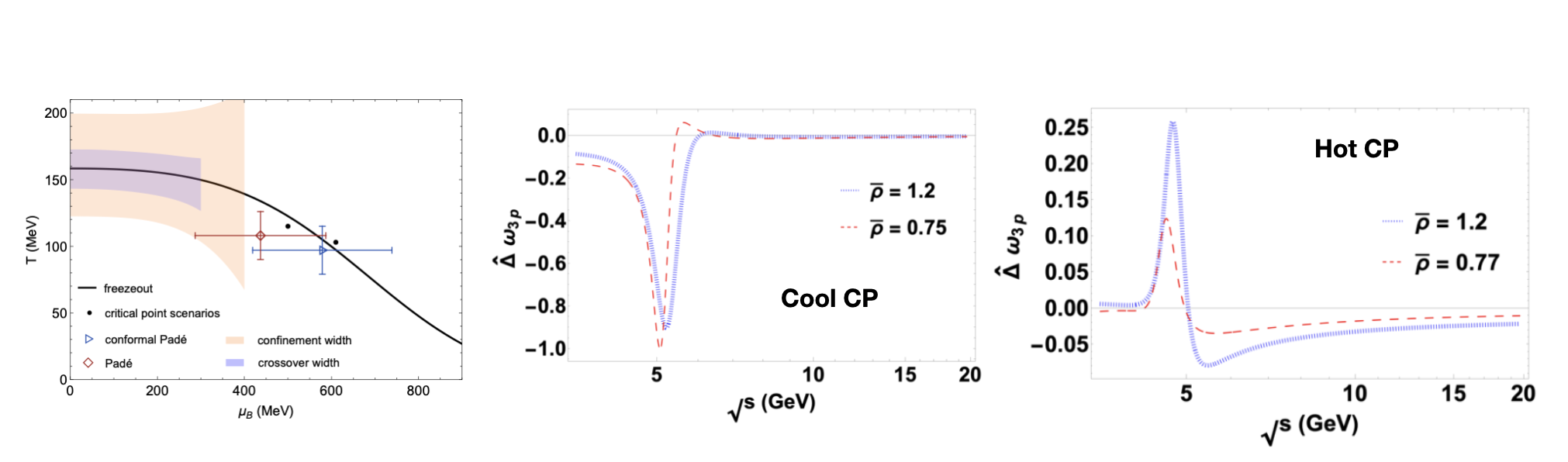}
 \caption{\textit{Left}: Estimated widths of confinement and chiral transitions from the lattice \cite{borsanyi2024qcddeconfinementtransitionline}.  The two estimates based on Pade resummation for the critical point are shown in red and blue. The black dots mark two representative critical-point locations, referred to as the \textit{cool} and \textit{hot} critical points, for which the freeze-out temperature at the critical chemical potential is above and below the critical temperature, respectively. \textit{Center and Right: } Normalized third factorial cumulant of the proton multiplicity as a function of collision energy for the cool (center) and hot (right) critical-point scenarios. The red and blue curves correspond to critical equations of state with   lower and upper $1\sigma$ bounds of $\bar{\rho}$ inferred from the Lee-Yang singularity analysis. The peak position is strongly constrained within the $1\sigma$ uncertainty of $\bar{\rho}$. }
    \label{fig:LYEcombined}
\end{figure}

Within the allowed parameter space, four distinct topological scenarios were identified, which are distinguished by the relative positions of the critical point, the crossover line, and the freeze-out curve. Each topology gives rise to experimentally distinguishable critical signatures as a function of collision energy, particularly for the third factorial cumulant, as illustrated in the center and right plots of Fig.~(\ref{fig:LYEcombined}). Of particular interest is the previously unexplored scenario corresponding to the case where the critical point lies below the freeze-out curve. Such a scenario cannot be excluded because the confinement-deconfinement crossover extends over a broader temperature range than the chiral crossover (see left plot of Fig.~(\ref{fig:LYEcombined})).

\subsection{Current developments}

\begin{figure}[h]
    \centering    \includegraphics[width=0.8\textwidth]{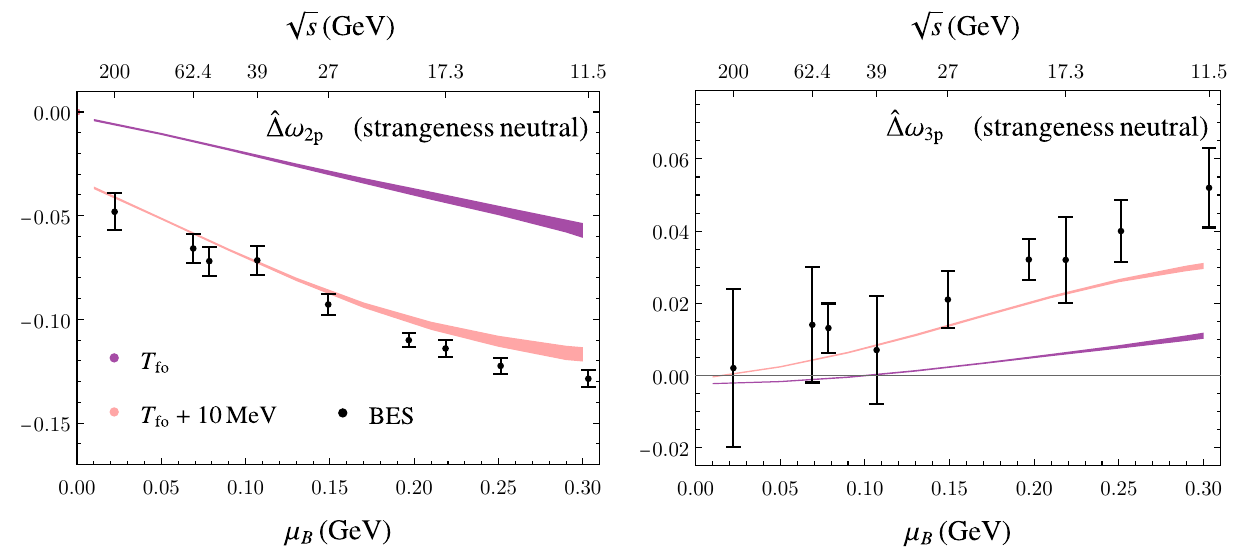}\hfill
    \caption{Factorial cumulants of the \emph{proton} (not net-proton) multiplicity of second (left) and third
(right) order: STAR beam-energy scan proton data from Ref.~\cite{STAR:2025zdq} compared with
theoretical predictions from the ongoing work \cite{PradeepBasarStephanov26}. 
Purple band : prediction obtained with the lattice QCD equation of state\cite{Abuali:2025tbd}, assuming
the hydrodynamic correlations have relaxed to their local-equilibrium values and applying
maximum-entropy freeze-out within the acceptance $|\Delta y|\leq 0.5$ and
$0.4~\mathrm{GeV}<p_T<2~\mathrm{GeV}$. Both the hydrodynamic cumulants and the mean
densities freeze out along the empirical freeze-out parametrization of
Ref.~\cite{Andronic:2017pug} with additional constraints of strangeness neutrality (mean strangeness density $n_s=0$ and ratio of mean electric charge density to mean baryon density, $n_Q/n_B=0.4$). Pink band : same as the purple band, but with the
hydrodynamic cumulants freezing out at a temperature $10\, \rm{MeV}$ above the freeze-out
temperature of the mean yields. The results do not test the presence of a critical point, but test the description of proton-number fluctuations in the region where lattice QCD is reliable.}
    \label{fig:current}
\end{figure}

The studies above model the critical contribution to the EoS through the Ising mapping. A complementary strategy is to apply maximum-entropy freeze-out directly to equations of state obtained using lattice QCD computations and to confront the results with BES measurements. At small values of chemical potentials or equivalently large collision energies, where the lattice extrapolations are controlled, these studies can be used to make predictions for experimentally observed factorial cumulants of proton multiplicities. Two recent efforts along these lines are described below.

A recent study presented at the SQM conference~\cite{PihanSQM} implemented a minimal
version of maximum-entropy freeze-out, in which only the baryon-density correlators are
matched, and computed proton multiplicity fluctuations using the 4D-TExS equation of state. The analysis assumed that hydrodynamic
fluctuations are locally equilibrated on an inhomogeneous freeze-out hypersurface obtained
from hydrodynamic simulations. The authors also presented exploratory Bayesian studies
aimed at inferring the baryon susceptibility as a function of collision energy by
combining BES measurements with their hybrid framework \cite{Pihan:2026xng}.

In an ongoing work~\cite{PradeepBasarStephanov26}, maximum-entropy freeze-out is used to map
lattice QCD susceptibilities onto proton multiplicity fluctuations in heavy-ion
collisions, over the range of baryon chemical potential where the Taylor expansion
remains under control. The analysis rests on three assumptions:
(i) the freeze-out hypersurface is homogeneous;
(ii) the normal to the hypersurface points along the local fluid velocity; and
(iii) hydrodynamic correlations have relaxed to their local-equilibrium values at
freeze-out.
The resulting second and third factorial cumulants are compared with BES measurements in Fig.~\ref{fig:current}. When the hydrodynamic cumulants and the mean densities freeze out at the same temperature (purple band), the equilibrium expectations differ quantitatively from the data. When the hydrodynamic cumulants instead freeze out above the freeze-out temperature of the mean yields (pink band), the equilibrium expectations agree better with the experimental data. This may be interpreted as a memory effect: correlation functions relax more slowly than the mean densities and therefore retain the imprint of an earlier, hotter stage of the evolution. A simple relaxational model for the evolution of out-of-equilibrium correlations can be used to estimate the difference between the freeze-out temperature of the second cumulant and that of the mean hydrodynamic densities; this difference turns out to be about 10 MeV~\cite{PradeepBasarStephanov26}.

Assumptions (ii) and (iii) are shared with Ref.~\cite{PihanSQM}, and neither work
simulates the out-of-equilibrium evolution of hydrodynamic correlations. The two analyses
differ in their remaining choices. Ref.~\cite{PihanSQM} uses an inhomogeneous freeze-out
hypersurface, matches the baryon-density correlators, and introduces dynamical effects by
enforcing global conservation laws through SAM~3.0~\cite{Poberezhniuk:2026bfv}.
Ref.~\cite{PradeepBasarStephanov26} uses a homogeneous hypersurface, matches by
construction all combinations of hydrodynamic cumulants of the energy, baryon,
electric-charge, and strangeness densities, and mimics dynamical memory effects by
freezing out the hydrodynamic cumulants at a higher temperature than the mean densities.

\section{Dynamical evolution of hydrodynamic fluctuations near the critical point}
\label{Sec:Dyn}
The freeze-out studies of Sec.~\ref{Sec:ME} take the hydrodynamic state at freeze-out as given. The final element of the pipeline is the dynamical evolution that produces this state, and the critical point influences both the average trajectory of the fireball and the evolution of its fluctuations. The presence of a critical point can significantly distort the hydrodynamic trajectories of the fireball. In the limit of weak dissipation, the hydrodynamic trajectories are represented by isentropes. The universal topography of the isentropes near the critical point results in the well-known focusing phenomenon, and leads to qualitatively distinct adiabatic expansion scenarios determined by the non-universal mapping parameters and entropy per baryon at the critical point \cite{Pradeep:2024cca}. One interesting scenario results in adiabatic trajectories entering and leaving the coexistence region via the same phase.

Turning to the fluctuations themselves, all of the studies discussed in the previous section relied on the assumption that hydrodynamic fluctuations remain close to local equilibrium throughout the evolution until freeze-out. This approximation eventually breaks down sufficiently close to the critical point because the relaxation time of the critical slow mode grows rapidly, 
leading to critical slowing down. 
Phenomenological studies  show that memory effects prevent the fluctuations from fully relaxing to their equilibrium values and can lead to enhanced Gaussian multiplicity fluctuations at freeze-out compared to equilibrium expectations\cite{Pradeep:2022mkf}.

 Describing such memory effects quantitatively, beyond the Gaussian level, requires dynamical equations for higher-order correlators. A major recent theoretical advance has been the derivation of deterministic equations governing the evolution of non-Gaussian fluctuation correlators of hydrodynamic densities \cite{An:2026glk}. 
 Another promising development is the application of the Metropolis algorithm to fluctuating hydrodynamics \cite{Bhambure:2024gnf}, which provides a promising alternative to Langevin-based approaches by naturally avoiding issues such as multiplicative noise and short distance singularities. An attractive feature of this framework is that dissipation emerges from the thermal fluctuations themselves, offering the potential for more robust and stable relativistic hydrodynamic simulations, although a full $3+1$ D implementation remains to be developed. For a comprehensive review of the recent developments in relativistic hydrodynamic fluctuations, see Ref.~\cite{Basar:2024srd}.
\section{Summary and looking forward}
\label{Sec:Sum}

Interpreting the experimental search for the QCD critical point requires a theoretical pipeline connecting the QCD equation of state to fluctuation observables (Fig.~\ref{Fig:pipeline}). Lattice QCD and functional methods provide bounds and estimates on the location of the critical point, and the EoS near it is beginning to be constrained. Maximum entropy freeze-out translates these constraints into proton factorial cumulants. 
Incorporating the out-of-equilibrium evolution of critical fluctuations, starting from a realistic initial state, remains the main open challenge; until then, comparisons with BES data such as Fig.~\ref{fig:current} should be viewed as indicative rather than system-specific. Implementing it, together with precision BES measurements, will be essential for determining whether, and where, QCD has a critical point.

\section{Acknowledgments}
The author is grateful to the organizers of SQM for giving the opportunity to present this talk. The author is supported via Ramanujan Fellowship, Project File Number RJF/2025/000614.



\bibliographystyle{elsarticle-num}
\bibliography{sqm2026_template}



\end{document}